\documentclass[aps, prb, 10pt, twocolumn, superscriptaddress, notitlepage, floatfix, longbibliography]{revtex4-2}
\usepackage{amsmath, amsfonts, amssymb, amsthm, mathtools, dsfont}
\newcommand\U{\textrm{U}(1)}
\usepackage{graphicx}
\usepackage{hyperref}
\usepackage{indentfirst}
\usepackage{microtype}

\usepackage{pgfplots}
\pgfplotsset{compat=1.15}
\usepackage{mathrsfs}
\usetikzlibrary{arrows}

\newcommand{\jkh}[1]{\left\langle #1 \right\rangle}

\newcommand{\fkh}[1]{\left[ #1 \right]}
\newcommand{\kh}[1]{\left( #1 \right)}
\def\bO{\mathcal{O}}
\newcommand{\euler}{\,\mathrm{e}}
\newcommand{\ii}{\mathrm{i}}
\newcommand*\diff{\mathop{}\!\mathrm{d}}
\newcommand*\Diff[1]{\mathop{}\!\mathrm{d^#1}}
\newcommand{\pdv}[1]{\frac{\partial}{\partial #1}}
\newcommand{\vb}[1]{\boldsymbol{#1}}
\newcommand{\vu}[1]{\hat{\boldsymbol{#1}}}

\DeclarePairedDelimiter\ket{\lvert}{\rangle}
\DeclarePairedDelimiterX\braket[2]{\langle}{\rangle}{#1\,\delimsize\vert\,\mathopen{}#2}
\DeclareMathOperator{\Tr}{Tr}
\DeclarePairedDelimiter\abs{\lvert}{\rvert}%
\DeclarePairedDelimiter\norm{\lVert}{\rVert}%
\makeatletter
\let\oldabs\abs
\def\abs{\@ifstar{\oldabs}{\oldabs*}}
\let\oldnorm\norm
\def\norm{\@ifstar{\oldnorm}{\oldnorm*}}
\makeatother

\begin{document}
\title{Disorder operators in two-dimensional Fermi and non-Fermi liquids through multidimensional bosonization}
\author{Kang-Le Cai}
\author{Meng Cheng}
\affiliation{Department of Physics, Yale University, New Haven, CT 06520-8120, USA}

\begin{abstract}
Disorder operators are a type of non-local observables for quantum many-body systems, measuring the fluctuations of symmetry charges inside a region. It has been shown that disorder operators can reveal global aspects of many-body states that are otherwise difficult to access through local measurements. We study the disorder operator for U(1) (charge or spin) symmetry in two-dimensional Fermi and non-Fermi liquid states, using the multidimensional bosonization formalism. For a region $A$, the logarithm of the charge disorder parameter in a Fermi liquid with  isotropic interactions scales asymptotically as $l_A\ln l_A$, with $l_A$ being the linear size of the region $A$. We calculate the proportionality coefficient in terms of Landau parameters of the Fermi liquid theory. We then study models of the Fermi surface coupled to gapless bosonic fields realizing non-Fermi liquid states. In a simple spinless model, where the fermion density is coupled to a critical scalar, we find that at the quantum critical point the scaling behavior of the charge disorder operators is drastically modified to $l_A \ln^2 l_A$. We also consider the composite Fermi liquid state and argue that the charge disorder operator scales as $l_A$. 
\end{abstract}
\maketitle

\section{Introduction}\label{Sec_INTRO}
For quantum many-body states with global symmetry, it has become increasingly evident that fully characterizing the global aspects of the states requires the knowledge of symmetry disorder operators (also known as defect operators in recent literature). A disorder operator $U_A(g)$ is defined as symmetry transformation $g$ restricted to a subregion $A$ of the system. On one hand, these operators have proven instrumental in the classification of gapped topological phases with global symmetry, which is based on the theoretical understanding of the formal structures of defect operators. Furthermore, the expectation value of the disorder operator $\jkh{U_A(g)}$, named disorder parameter, provides a new class of non-local observables for quantum many-body systems, going beyond conventional probes based on local operators. For a gapped symmetry-preserving ground state, the disorder parameter decays exponentially with the volume of the boundary $\partial A$:
\begin{align}
    \ln \abs{\jkh{U_A(g)}}=-\alpha \abs{\partial A}+\gamma_g+\ldots,\label{eq:Uscaling}
\end{align}
where $\alpha$ is a non-universal parameter. This area-law behavior is similar to that of entanglement entropy, while the subleading correction $\gamma_g$ is a new type of universal many-body invariant. It has been shown that $\gamma_g$ is related to the ``quantum dimension" of the corresponding symmetry defect~\cite{chen2022topological, cai2025universal}. Similar scaling of $\jkh{U_A(g)}$ is also found in quantum critical states described by conformal field theories (CFTs), where $\gamma_g$ now depends on the shape of the region $A$. In the case where $A$ is a disk, $\gamma_g$ is referred to as the defect entropy and has been shown to be a monotone under the renormalization group~\cite{Cuomo2022}. When $A$ contains sharp corners, for a CFT state in two dimensions, the subleading correction takes the form of $s\ln l_A$, where $l_A$ is the linear size of $A$ and $s$ is a universal quantity for the CFT~\cite{Wu2021a, Wu2021b, Z2DisorderOp, U1DisorderOp, DQCPDisorderOp}. The logarithmic correction has been investigated in several important examples of two-dimensional (2D) lattice models for quantum criticality~\cite{DQCPDisorderOp, FermionCFTDO, liu2023disorder}, bringing new insights into the nature of the critical states.

While quite ubiquitous, well-known exceptions to the scaling form in Eq.~\eqref{eq:Uscaling} are found in compressible systems, the most familiar example being the Fermi liquid (FL). For such states and $g$ being a U(1) symmetry generated by charge or spin, the leading term receives a multiplicative logarithmic correction, i.e. $l_A^{d-1} \ln l_A$ in $d$ spatial dimensions, with the coefficient determined by the geometric shapes of the Fermi surface and the region $A$. For non-interacting fermions the precise form of the scaling has been determined for charge fluctuations~\cite{song2012bipartite, FermionDisorderOp, Swingle2012} and entanglement entropy (both von Neumann and R\'enyi)~\cite{Wolf:2006zzb, Gioev2006, FermionDisorderOp}.

In this work, we focus on the study of the disorder operator for U(1) charge conservation symmetry in 2D interacting FLs and non-Fermi liquids (NFLs). Computing the disorder parameter and related entanglement measures analytically or numerically in an interacting many-body system is typically challenging, particularly in spatial dimensions higher than 1. To circumvent the difficulty, we employ the multidimensional bosonization technique~\cite{haldane2005luttingers, Houghton:1992dz, neto1994bosonization,neto1995exact}. In the bosonization formalism, the fermion density operator is linear in the boson basis, rendering calculations related to fermion density fluctuations considerably easier. It has been used to compute the disorder parameter for 2D free fermions~\cite{tan2020particle} and the entanglement entropy for interacting 2D fermion systems~\cite{ding2012entanglement}. Within the bosonization framework, we will show that the U(1) charge disorder operator is essentially determined by the charge structure factor, which will then be computed in several prototypical examples in two dimensions. 
 
Our main results are the following.

(1) For the Landau FL with a circular Fermi surface, $\ln \abs{\jkh{U_A}}$ scales as $l_A \ln l_A$, and we determine the coefficient exactly in terms of Landau parameters for isotropic interactions.

(2) We study the Fermi surface with the number density coupled to a gapless scalar, as a model for the NFL. We show that the scaling of $\ln \abs{\jkh{U_A}}$ is dramatically modified to $l_A \ln^2 l_A$. We also consider the example of the composite Fermi liquid and argue that the charge disorder operator obeys the $\sim l_A$ scaling. We will show that the theoretical results can naturally explain the main qualitative features observed in a recent numerical study~\cite{FermionDisorderOp} on charge and spin disorder operators in a NFL at a ferromagnetic quantum critical point~\cite{NFLFM}.

The paper is organized as follows. We first discuss the general aspects in Sec.~\ref{Sec_DO}. Then we review the operator formalism of multidimensional bosonization in Sec.~\ref{Sec_MB}. In particular, we fix a problem regarding the diagonalization of the bosonized Hamiltonian using Bogoliubov transformations in the original literature~\cite{neto1995exact}. In Sec.~\ref{Sec_BC}, we apply the method to 2D FLs, re-deriving the results for free fermions~\cite{tan2020particle} and then extend the results to the model with isotropic density-density interactions. In Sec.~\ref{Sec_NFL}, we study two models of NFLs: one in which spinless fermions are coupled to a gapless free scalar, and the other being the celebrated composite Fermi liquid. In Sec.~\ref{Sec_discussions}, we compare the theory with a recent numerical study~\cite{FermionDisorderOp}.

\section{Disorder operator in 2d Fermi gas}\label{Sec_DO}
The disorder operator of U(1) fermion number conservation symmetry in region $A$ is defined as $\euler^{\ii\theta N_A}$, where $\theta$ is a real parameter and $N_A$ is the total fermion number operator in region $A$. The ground state expectation 
\begin{equation}
    Z(\theta)=\jkh{\euler^{\ii\theta N_A}}  
\end{equation}
is also named the particle number cumulant generating functional or full counting statistics in the literature~\cite{FCS1, FCS2, FCS3}, since it generates cumulants of particle number distribution
\begin{align}
    \jkh{N_A^n}_c = \kh{-\ii\pdv{\theta}}^n \ln Z(\theta) \biggr\rvert_{\theta=0}.
\end{align}
We can use the cumulants to expand $Z(\theta)$ as
\begin{align}
    \ln Z(\theta) = \sum_{n=0}^{\infty} \frac{(\ii\theta)^n}{n!} \jkh{N_A^n}_c.\label{DO_Eq_Cumulant}
\end{align}
The cumulants can also be used to expand the entanglement entropy and R\'enyi entropy~\cite{song2012bipartite} for non-interacting fermions, while the expansion in Eq.~\eqref{DO_Eq_Cumulant} is true for interacting fermions in general.

For a large enough region, the cumulants for non-interacting fermions in $d$ spatial dimensions are known to scale like~\cite{tam2022topological}
\begin{align}
    \jkh{N_A^k}_c \sim (k_F R)^{d+1-k},\label{DO_Eq_NAScale}
\end{align}
where $R$ is the length scale for region $A$. Thus the most divergent terms have $k\leq d+1$. For $d=2$ here, the only relevant terms are
\begin{align}
    Z(\theta) \simeq \exp\kh{-\frac{1}{2} \theta^2 \jkh{N_A^2}_c - \frac{\ii}{6} \theta^3 \jkh{N_A^3}_c},
\end{align}
where the $\jkh{N_A^3}$ term enters as a pure phase factor. Thus, in a Fermi gas, the density fluctuations obey an approximate Gaussian distribution.

To calculate $\jkh{N_A^2}_c$, we start from the (equal-time) connected density-density correlation (structure factor) $G(\vb{q})=\frac{1}{V}\jkh{\rho(\vb{q})\rho(\vb{-q})}_c$ and then use the following relation
\begin{align}
    \jkh{N_A^2}_c &=\int_{A} \diff{\vb{r}}\diff{\vb{r'}} \frac{1}{V^2} \sum_{\vb{q}} \jkh{\rho(\vb{q})\rho(-\vb{q})}_c \euler^{\ii \vb{q}\cdot(\vb{r}-\vb{r}')}\\
    &=\int_{A} \diff{\vb{r}}\diff{\vb{r'}} \int \frac{\diff^2 \vb{q}}{(2\pi)^2}
    G(\vb{q})\euler^{\ii \vb{q}\cdot(\vb{r}-\vb{r}')}.
    \label{DO_Eq_NAbyInt}
\end{align}

Assume that the Fermi surface is isotropic, so that $G(\vb{q})$ is independent of the direction of $\vb{q}$. We can evaluate the integral Eq.~\eqref{DO_Eq_NAbyInt} as
\begin{align}
    \jkh{N_A^2}_c&=\frac{\pi R^4}{2} \int_{0}^{\frac{1}{\alpha}} q G(q) \left[{}_{0}F_1\kh{2,-\frac{q^2 R^2}{4}}\right]^2\diff q \notag\\
    &=2\pi R^2 \int_{0}^{\frac{1}{\alpha}} \frac{G(q)}{q}J_1^2(qR)\diff q\notag\\
    &=2\pi R \int_{0}^{\frac{R}{\alpha}} \frac{G(x/R)}{x/R}J_1^2(x)\diff x,\label{DO_Eq_NAbyInt1}
\end{align}
where $J_1$ is the Bessel function of the first kind, and we have introduced a high-$q$ cut-off $\frac{1}{\alpha}$. The leading $R$ dependence is determined by the behavior of $G(q)$ as $q\rightarrow 0$. Let us assume $G(q)\sim a q^n$, with $n\geq 0$. The result of the integral to the leading order in $R$ is given by
\begin{align}
    \jkh{N_A^2}_c=\begin{cases}
        a\pi R^2+\cdots & n=0\\
        2aR\ln\frac{R}{\alpha}+\cdots & n=1\\
        \frac{2a}{n-1}\alpha^{1-n}R+\cdots & n\geq 2
    \end{cases}.\label{DO_Eq_NA}
\end{align}
Details of the evaluation of the integral Eq.~\eqref{DO_Eq_NAbyInt1} can be found in Appendix~\ref{integral}.

It is also interesting to consider subleading corrections in $\langle N_A^2\rangle$, which are known to depend on the shape of the region $A$. More specifically, the form of the subleading correction depends on the opening angles of the sharp corners of $A$, denoted by $\alpha_i$. Let us discuss the subleading corner correction case by case.

For $n\geq 2$, the subleading correction is a constant $b(\alpha)$, which was shown in Ref.~\cite{Estienne:2021hbx} to be a universal function of $\alpha$:
\begin{equation}
    b(\alpha)=-(1+(\pi-\alpha)\cot\alpha)C,
\end{equation}
where $C=a$ for $n=2$, and $0$ if $n>2$. 

For $n=1$, the subleading term is in general proportional to $R$. The precise form of the correction can be found in Ref.~\cite{Estienne:2021hbx}.

Finally, we discuss the emergence of cusp singularities in $Z(\theta)$. In lattice systems, $Z(\theta)$ is manifestly $2\pi$ periodic in $\theta$, and the Gaussian-like dependence $\euler^{-\textrm{const}\cdot\theta^2}$ holds only within the interval $\theta\in(-\pi,\pi)$. Consequently, $-\ln Z(\theta)$ develops a cusp at $\theta=\pi$. For a finite region $M$, this cusp is always smoothed out close to $\theta=\pi$; however, in the limit where the size of $M$ tends to infinity—after taking the thermodynamic limit of the full system—the cusp becomes a true singularity, reflecting the quantization of charges. A similar phenomenon was highlighted in Ref.~\cite{WangFCS} for superfluid states. Related cusp singularities have also been observed in mixed states of one-dimensional (1D) spin chains\cite{Zang, cai2025universal}.

\begin{figure}[ht]
    \centering\includegraphics[width=\columnwidth]{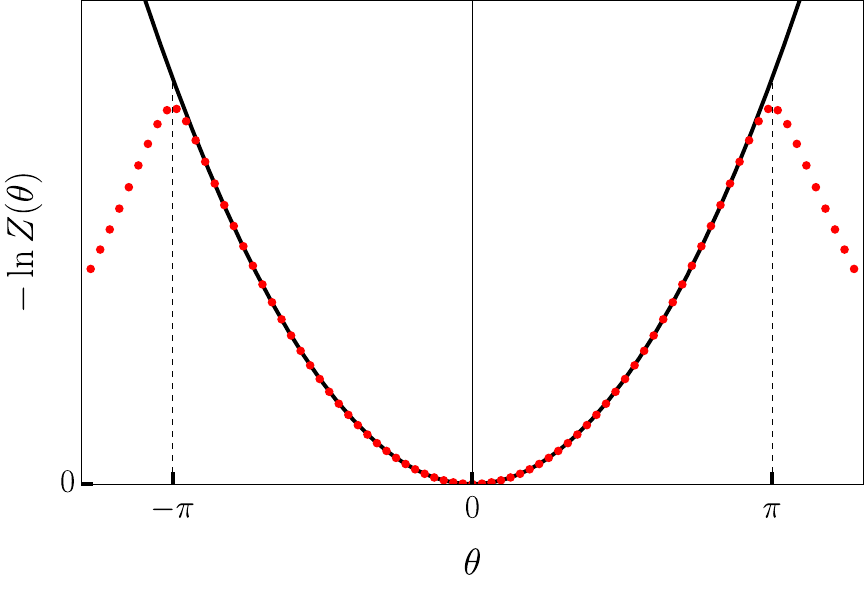}
    \caption{$-\ln Z(\theta)$ of a $50\times 50$ free fermion lattice model with a fixed $20\times 20$ region. The numerical results are fitted with a quadratic function $a\theta^2$ in the region $\theta\in (-\pi,\pi)$. $-\ln Z(\theta) \propto \theta^2$ is valid inside the region $\theta\in (-\pi,\pi)$ to a very good precision, and there is a smooth transition linking different intervals at $\theta=-\pi,\pi$.}\label{Fig_Z_Free}
\end{figure}

We plot the numerical results of $-\ln Z(\theta)$ in Fig.~\ref{Fig_Z_Free} for a fixed square region, in a model of free fermions hopping on a square lattice: 
\begin{align}
    H=-t\sum_{\langle ij \rangle} (c^{\dagger}_i c_j+\textrm{h.c.}).
\end{align}
Indeed, one clearly observes the cusp-like feature at $\theta=\pi$, which is rounded off by the finite-size effect.

\section{Multidimensional bosonization}\label{Sec_MB}
\subsection{Framework}
Multidimensional bosonization is a direct generalization of the well-known 1D bosonization. Intuitively, the physical picture for multidimensional bosonization is the following: at each point of the Fermi surface and for each excitation momentum $\vb{q}$ pointing outside, there is a bosonic mode similar to one branch of the Tomonaga-Luttinger liquid. At the Landau Fermi liquid fixed point, one can ignore scattering processes that change the fermion number at a given point of the Fermi surface, and only the forward scattering is kept.

We now review multidimensional bosonization in the operator formalism~\cite{neto1994bosonization,neto1995exact}. Suppose $c_{\vb{k}}$ and $c^{\dagger}_{\vb{k}}$ are the fermion annihilation and creation operators. Let the single-particle spectrum be $\epsilon(\vb{k})$. The particle velocity is defined as $\vb{v_{k}}=\nabla_{\vb{k}} \epsilon(\vb{k})$.

Consider the following particle-hole operator:
\begin{align}
    n_{\vb{q}}(\vb{k})=c_{\vb{k}-\frac{\vb{q}}{2}}^{\dagger} c_{\vb{k}+\frac{\vb{q}}{2}},
\end{align}
Similar to the 1D case, for each $\vb{k}_F$ on the Fermi surface and $\vb{q}\cdot \vb{v}_{\vb{k}_F}>0$, we can define the boson operators as
\begin{align}
    \begin{aligned}
        d_{\vb{q}}(\vb{k}_F)&=\sum_{\vb{k}} \Phi_{\Lambda}\kh{\abs{\vb{k}-\vb{k}_F}}n_{\vb{q}}(\vb{k}),\\
        d_{\vb{q}}^{\dagger}(\vb{k}_F)&=\sum_{\vb{k}} \Phi_{\Lambda}\kh{\abs{\vb{k}-\vb{k}_F}}n_{-\vb{q}}(\vb{k}),
    \end{aligned}
\end{align}
where $\Phi_{\Lambda}\kh{\abs{\vb{k}-\vb{k}_F}}$ is a dimensionless smearing function depending on the momentum cutoff $\Lambda$ that keeps $\vb{k}$ close to $\vb{k}_F$. By introducing $\Phi_{\Lambda}$, we are using spheres of radius $\Lambda$ to cover all the states near the Fermi surface. The approximation is valid when states of interest have momentum close to $\vb{k}_F$ with fluctuation $\vb{q}$ such that $q\ll \Lambda\ll k_F$. The condition $\vb{q}\cdot \vb{v}_{\vb{k}_F}>0$ is required to normal order the boson operators so that they annihilate the filled Fermi sea $\ket{\textrm{FS}}$:
\begin{align}
    d_{\vb{q}}(\vb{k}_F) \ket{\textrm{FS}}=0.
\end{align}
We will assume the condition $\vb{q}\cdot \vb{v}_{\vb{k}_F}>0$ for every boson operator at Fermi point $\vb{k}_F$ with excitation momentum $\vb{q}$ later in this paper.

In the long wavelength limit $(\vb{q}\rightarrow 0)$, we have the following commutation relation:
\begin{align}
    [n_{\vb{q}}(\vb{k}),n_{-\vb{q}'}(\vb{k}')]=\delta_{\vb{k},\vb{k}^{\prime}}\delta_{\vb{q},\vb{q}'}\vb{q}\cdot\vb{v_k} \delta(\mu-\epsilon(\vb{k})),
\end{align}
where $\mu=\epsilon(\vb{k}_F)$ is the chemical potential. Thus, the boson operators satisfy the commutation relations
\begin{align}
    \left[ d_{\vb{q}}(\vb{k}_F),d^{\dagger}_{\vb{q}'}(\vb{k}_F') \right] = N_{\Lambda}(\vb{k}_F) V \vb{q}\cdot \vb{v}_{\vb{k}_F}
    \delta_{\vb{k}_F,\vb{k}_F'}\delta_{\vb{q},\vb{q}'},
\end{align}
where $N_{\Lambda}(\vb{k}_F)$ is the local density of states
\begin{align}
    N_{\Lambda}(\vb{k}_F)=\frac{1}{V} \sum_{\vb{k}} \abs{\Phi_{\Lambda}(\abs{\vb{k}-\vb{k}_F})}^2 \delta(\mu-\epsilon(\vb{k}))
\end{align}
and $V$ is the volume of the system.

For simplicity, we rescale the boson operators to absorb the additional factors
\begin{align}
    b_{\vb{q}}(\vb{k}_F)=(N_{\Lambda}(\vb{k}_F)V \vb{q}\cdot \vb{v}_{\vb{k}_F})^{-\frac{1}{2}} d_{\vb{q}}(\vb{k}_F).
\end{align}
The new operators $b_{\vb{q}}(\vb{k}_F)$ satisfy the standard bosonic commutation relations:
\begin{align}
    \left[ b_{\vb{q}}(\vb{k}_F),b^{\dagger}_{\vb{q}'}(\vb{k}_F') \right] =\delta_{\vb{k}_F,\vb{k}_F'}\delta_{\vb{q},\vb{q}'}.
\end{align}

In the bosonization formalism, the Fourier component of the fermion density near the Fermi surface can be represented by
\begin{align}
    \rho(\vb{q})&=\sum_{\vb{k}} n_{\vb{q}}(\vb{k})\notag\\ 
    &= \sqrt{N_{\Lambda} V\vb{q}\cdot\vb{v}_{\vb{k}_F}}\sum_{\vb{k}_F}\kh{b_{\vb{q}}(\vb{k}_F)+b^{\dagger}_{-\vb{q}}(-\vb{k}_F)}.\label{MB_Eq_Density}
\end{align}
As in the 1D case, the fermion density operator is linear in the boson basis, making calculations related to fermion density fluctuation considerably easier through bosonization.

\subsection{Hamiltonian}
In the restricted Hilbert space of low-energy states, the free fermion Hamiltonian can be expressed by 
\begin{align}
    H_0=&\sum_{\vb{k}} \epsilon(\vb{k})c^{\dagger}_{\vb{k}} c_{\vb{k}}\notag\\
    =&\sum_{\vb{q},\vb{k}_F} \vb{q}\cdot \vb{v}_{\vb{k}_F} b^{\dagger}_{\vb{q}}(\vb{k}_F) b_{\vb{q}}(\vb{k}_F),
\end{align}
which is also diagonal in the boson basis.

Consider a general fermion density-density interaction:
\begin{align}
    H_{\textrm{int}}&=\frac{1}{2V}\sum_{\vb{p},\vb{p}^{\prime},\vb{q}}U_{\vb{p},\vb{p}^{\prime}}(\vb{q})
    n_{-\vb{q}}(\vb{p}) n_{\vb{q}}(\vb{p}').\label{MB_Eq_Interaction}
\end{align}
We assume its effect is only important near the Fermi surface, and then it can be expanded in the boson basis. From now on, we will also assume the Fermi surface and the interaction are isotropic, i.e., $\epsilon(\vb{k})\equiv \epsilon(k)$ and $U(\vb{q})\equiv U(q)$, where $q=\abs{\vb{q}}$. Then, we have
\begin{align}
    H_{\textrm{int}}=&\sum_{\vb{k}_F,\vb{k}_F',\vb{q}} \sqrt{(\vb{q}\cdot \vb{v}_{\vb{k}_F}) (\vb{q}\cdot \vb{v}_{\vb{k}_F'})} N_{\Lambda} U(q)\notag\\
    &\left[b^{\dagger}_{\vb{q}}(\vb{k}_F) b_{\vb{q}}(\vb{k}_F')+b_{\vb{q}}(\vb{k}_F)b_{-\vb{q}}(-\vb{k}_F')+ \textrm{h.c.} \right].\label{MB_Eq_Hint}
\end{align}
Since the fermion density is linear in the boson operators, the interacting Hamiltonian is equivalent to a quadratic bosonic theory, which can be diagonalized by Bogoliubov transformation.

For numerical calculation, we will divide the Fermi surface into $N$ patches with one Fermi vector $\vb{k}_F$ chosen for each patch. At the end of the day, we will take the $N\rightarrow\infty$ limit so that finite-size effects do not contribute. The local density of states is
\begin{align}
    N_{\Lambda}=\frac{N(0)}{N},
\end{align}
where $N(0)$ is the total density of states at the Fermi surface. $N(0)=\frac{k_F}{2\pi v_F}$ for a 2D system.

The dimensionless coupling constant is defined as 
\begin{align}
    g=N(0) U(q).
\end{align}
Since the interaction is isotropic, $g$ is closely related to the dimensionless isotropic Landau parameter $F_0$ in Landau's Fermi liquid theory. In the following, we will show that $F_0=2g$.

\subsection{Explicit diagonalization in two dimensions}
\begin{figure}[htbp]
    \centering
\begin{tikzpicture}[line cap=round,line join=round,x=2cm,y=2cm]
\begin{axis}[ticks=none,x=2cm,y=2cm,axis lines=middle,
xmin=-1.5,xmax=1.5,ymin=-1.5,ymax=1.5]
\clip(-2.3242996205962068,-1.6976712074676128) rectangle (2.6867049683830198,2.1339415356820015);
\draw [line width=1.2pt] (0,0) circle (2cm);
\draw [-stealth,line width=1.2pt] (0,0) -- (0,0.5);
\draw [-stealth,line width=1.2pt] (0.7071067811865476,0.7071067811865476) -- (0.7071067811865476,1.2071067811865475);
\draw [-stealth,line width=1.2pt] (0,0) -- (0.7071067811865476,0.7071067811865476);
\draw [-stealth,line width=1.2pt] (-0.7071067811865476,-0.7071067811865476) -- (-0.7071067811865476,-1.2071067811865475);
\draw [-stealth,line width=1.2pt] (0,0) -- (-0.7071067811865476,-0.7071067811865476);
\begin{scriptsize}
\draw[color=black] (-0.15,0.28) node {\large $\vb{q}$};
\draw[color=black] (0.88,1.1) node {\large $b_{j}$};
\draw[color=black] (0.4,0.64) node {\large $\vb{k}_{j}$};
\draw[color=black] (-0.88,-1.1) node {\large $a_{j}$};
\draw[color=black] (-0.4,-0.64) node {\large $-\vb{k}_{j}$};
\end{scriptsize}\end{axis}\end{tikzpicture}
\caption{Illustration of the modes defined by Eq.~\eqref{MB_Eq_ab}.}
\label{Fig_FS_Mode}
\end{figure}
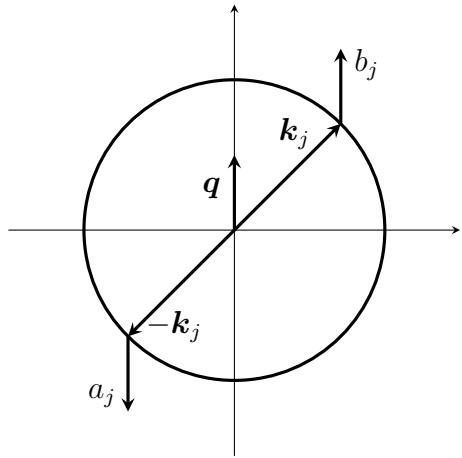

In this section, we specialize to 2D systems. For each $\vb{q}$, we can label the patches of the Fermi surface by the angle $\theta_j=\frac{2\pi j}{N}$, with $j=0$ in the direction of $\vb{q}$. To satisfy the condition $\vb{q}\cdot \vb{v}_{\vb{k}_F}=qv_F\cos\theta_j>0$, the allowed values of $j$ are $-\frac{N}{4}+1\leq j \leq \frac{N}{4}-1$. This restriction on indices is assumed for every summation afterwards.

The interaction Eq.~\eqref{MB_Eq_Hint} only couples $+\vb{q}$ modes with $-\vb{q}$ modes. As shown in Fig.~\ref{Fig_FS_Mode}, for each $\vb{q}$, we introduce the following notation:
\begin{align}
    \begin{aligned}
        b_{\vb{q}}(\vb{k}_j)&=b_j,\\
        b_{-\vb{q}}(-\vb{k}_j)&=a_j.\label{MB_Eq_ab}
    \end{aligned}
\end{align}
Then, the Hamiltonian can be rewritten as
\begin{align}
    H=\frac12 \sum_{\vb{q}} q v_F \mathcal{H}(q),
\end{align}
with
\begin{align}
    \mathcal{H}(q)=&\sum_{i} s_i\kh{b_i^{\dagger}b_i+a_i^{\dagger}a_i}\notag\\
    &+2\tilde{g}\sum_{i,j}\sqrt{s_i s_j}\kh{b_i^{\dagger}b_j+a_i^{\dagger}a_j+a_i b_j+b_i^{\dagger}a_j^{\dagger}},\label{MB_Eq_Hq}
\end{align}
where $s_j$ is defined as $s_j=\cos\frac{2\pi j}{N}$ and $\tilde{g}=\frac{g}{N}$. 

$\mathcal{H}(q)$ is a standard quadratic Hamiltonian that can be diagonalized using numerical methods. In Appendix~\ref{App_Diag}, we discuss the diagonalization of the Hamiltonian using Bogoliubov transformations. We find that the Bogoliubov transformation used in Ref.~\cite{neto1995exact} is actually singular, and a non-singular one can be found once taking into account the symmetry $i\leftrightarrow -i$. 

We also analytically solve for the collective mode in the $N\rightarrow \infty$ limit. We find a collective mode with the following dispersion relation:
\begin{align}
    \omega=q v_F \frac{1+2g}{\sqrt{1+4g}}.
\end{align}
This is precisely the zero sound of the Fermi surface, which can also be seen from the semiclassical Boltzmann equation in Landau Fermi liquid theory. In Appendix~\ref{App_ZeroSound}, we find the sound velocity in two dimensions to be
\begin{align}
    v_0= v_F \frac{1+F_0}{\sqrt{1+2F_0}}.
\end{align}
Comparing the two results, we have the relation $F_0=2g$.

\section{Disorder operator in Fermi liquid from bosonization}\label{Sec_BC}
In this section, we apply the bosonization formalism to compute the disorder operator $Z(\theta)$ in Fermi liquid states.

As noted in Sec.~\ref{Sec_DO}, the disorder operator for non-interacting fermions is completely determined by $\jkh{N_A^2}_c$. The bosonization framework capture essentially the leading second-order cumulant in the series Eq.~\eqref{DO_Eq_Cumulant}. There is no third-order cumulant because the three-point function of the density operator is automatically zero in the bosonization formalism. Similarly, there are no higher-order connected correlations. Within bosonization, we have
\begin{align}
    Z(\theta) = \exp\kh{-\frac{1}{2} \theta^2 \jkh{N_A^2}_c}.
\end{align}
Therefore, the problem reduces to the computation of $\jkh{N_A^2}_c$, and the general discussion in Sec.~\ref{Sec_DO} (especially  Eq.~\eqref{DO_Eq_NA}) still applies. We will thus focus on the computation of the structure factor $G(q)$.

Meanwhile, it is worth noting that odd-order cumulants do not contribute to entanglement entropy and R\'enyi entropy~\cite{song2012bipartite} for non-interacting fermions. However, they do contribute to tripartite mutual information.

\subsection{Non-interacting fermions}
We first consider 2D non-interacting fermions. The structure factor can be found using Wick's theorem:
\begin{align}
    G_0(q) = \int \frac{\diff^2{\vb{k}}}{(2\pi)^2}\, f(\vb{k})(1-f(\vb{k}+\vb{q})),\label{BC_Eq_G0}
\end{align}
where $f(\vb{k})=\theta(E_F-\epsilon(\vb{k}))$ is the Fermi distribution function. The above integral has the meaning of measuring the volume of the Fermi sea that is moved outside after shifting by $\vb{q}$. For a circular Fermi surface and $q<2k_F$, the integral evaluates to
\begin{align}
    G_0(\vb{q})&=\frac{1}{(2\pi)^2} \left[ \kh{\pi-2\arccos{\frac{q}{2k_F}}} k_F^2 +q \sqrt{k_F^2 - \frac{q^2}{4}} \right]\notag\\
    &= \frac{k_F^2}{2\pi^2} \frac{q}{k_F} +\bO(q^3).
\end{align}
Generally, when $q$ is small relative to the length scales of the Fermi surface, Eq.~\eqref{BC_Eq_G0} can be approximated to first order in $q$ by an integral on the Fermi surface:
\begin{align}
    G_0(\vb{q}) = \int_{\textrm{FS}} \theta(\vb{q}\cdot\vu{n}) \vb{q}\cdot\vu{n}\frac{\diff{S}}{(2\pi)^2},
\end{align}
where $\vu{n}$ is the unit normal vector.

Using bosonization, we can reproduce this leading $q$ dependence of $G_0(q)$ as
\begin{align}
    G_0(q)=\sum_{\vb{k}_F} N_{\Lambda}\vb{q}\cdot\vb{v}_{\vb{k}_F}
    =\frac{k_F}{2\pi^2}q.
\end{align}
By Eq.~\eqref{DO_Eq_NA}, we have
\begin{align}
    Z(\theta) \simeq \exp\kh{ -\theta^2 \frac{k_F}{2\pi^2} R\ln\frac{R}{\alpha} }.
\end{align}

\subsection{Interacting fermions}\label{Sec_BC_IntF}
For interacting fermion systems, we need to use the diagonalized modes in the fermion density Eq.~\eqref{MB_Eq_Density}.

\begin{figure}[t]
    \centering
    \includegraphics[width=\columnwidth]{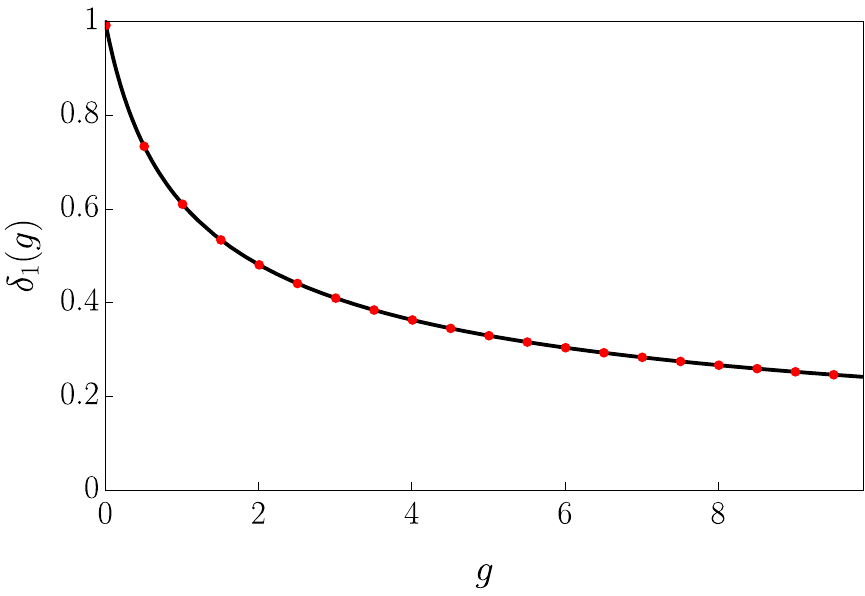}
    \caption{The ratio $\delta_1(g)$ vs contact interaction strength $g$. Numerical results with $N=320$ are shown in red dots, while the solid black line is the field theory prediction Eq.~\eqref{BC_Eq_deltag}.}
    \label{Fig_Ratio_Int_ZeroT}
\end{figure}

While $G(q)$ remains linear in $q$, numerical diagonalization shows that a repulsive interaction tends to reduce the slope of $G(q)$, thereby decreasing the scaling coefficient of $Z(\theta)$. This is consistent with corresponding results for 1D Luttinger liquid~\cite{song2012bipartite}. We define the ratio as
\begin{align}
    \delta_1(g)=\frac{G(q)}{G_0(q)}.
\end{align}
Then, the disorder parameter becomes
\begin{align}
    Z(\theta) \simeq \exp\kh{ -\theta^2 \frac{k_F}{2\pi^2} \delta_1(g) R\ln\frac{R}{\alpha} }.
\end{align}
In Fig.~\ref{Fig_Ratio_Int_ZeroT}, we plot numerical results of $\delta_1(g)$ with $N=320$.

Alternatively, $G(q)$ can be calculated using field-theoretic bosonization with Hubbard-Stratonovich transformation~\cite{houghton2000multidimensional, swingle2013universal}. We find 
\begin{align}
    \ii G(\omega,q)=\frac{\ii G_0(\omega,q)}{1+2U(q)\ii G_0(\omega,q)},\label{BC_Eq_G}
\end{align}
which is identical to the random phase approximation (RPA) result. Here $G_0(\omega,q)$ is the full frequency-dependent density-density correlation for bosonized free fermions:
\begin{align}
    \ii G_0(\omega,q)=\frac{k_F}{2\pi v_F}\kh{1-\frac{\abs{\omega}}{\sqrt{\omega^2-v_F^2 q^2}}}.
\end{align}
Integrating over $\omega$ in Eq.~\eqref{BC_Eq_G}, we end up with~\footnote{The integral is evaluated for $-\frac{1}{2}<g<0$, where no divergence occurs, and is then analytically continued to $g>0$.}
\begin{align}
    \delta_1(g)=\frac{\sqrt{1+4g}+4g \arctan\sqrt{1+4g}}{(1+4g)^{\frac32}}.\label{BC_Eq_deltag}
\end{align}

For large $g$, the dominant contribution to $G(q)$ comes from the collective mode pole~\cite{swingle2013universal}, and $\delta_1(g)\sim \frac{\pi}{4\sqrt{g}}$.

\subsection{Finite temperature}\label{Sec_FT}
In this section, we consider thermal fluctuations of the disorder operator $Z(\theta,\beta)=\Tr \euler^{-\beta H} \euler^{\ii\theta N_A}$.

For non-interacting fermions, we can calculate $G_0(q)$ at finite temperature using bosonization as
\begin{align}
    G_0(q)=\frac{k_F}{(2\pi)^2}q\int_{-\frac{\pi}{2}}^{\frac{\pi}{2}} \cos x\coth\frac{\beta q v_F \cos x}{2}\diff x.
\end{align}
As $q\rightarrow 0$, $G_0(q)$ approaches a constant value:
\begin{align}
    G_0(0)\simeq \frac{k_F}{2\pi\beta v_F},
\end{align}
which can also be derived using Wick's theorem. Using Eq.~\eqref{DO_Eq_NA}, we see that
\begin{align}
    Z(\theta,\beta)\simeq \exp\kh{-\theta^2 \frac{k_F}{4\beta v_F} R^2},
\end{align}
indicating that the scaling of $Z(\theta,\beta)$ exhibits volume-law behavior at any finite temperature.

If we consider the interacting Fermi liquid model, we have
\begin{align}
    G(0)\simeq \frac{k_F}{2\pi\beta v_F} \delta_2(g),
\end{align}
where $\delta_2(g)$ is another dimensionless ratio that depends only on $g$. 

\begin{figure}[t]
    \centering
    \includegraphics[width=\columnwidth]{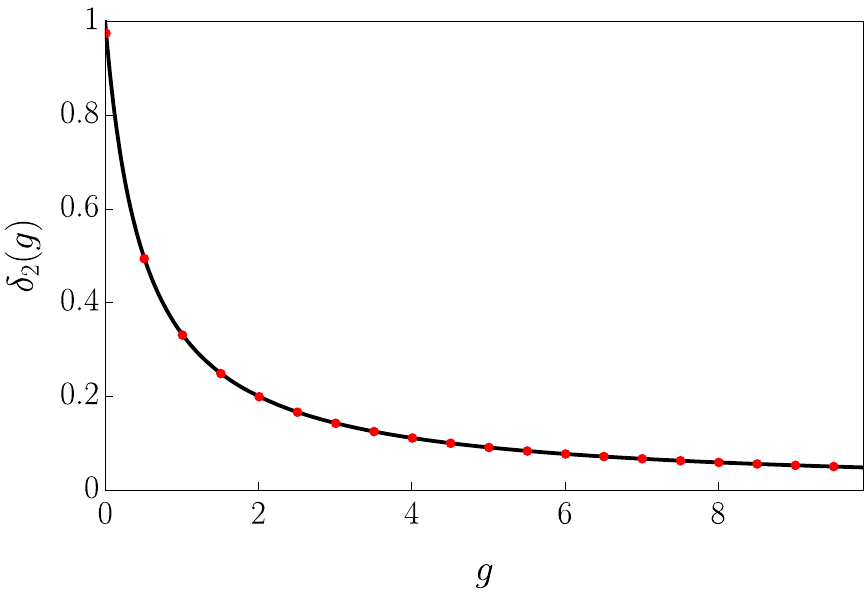}
    \caption{The ratio $\delta_2(g)$ vs contact interaction strength $g$. Numerical results with $N=320$ are shown in red dots and the solid black line is the function $\frac{1}{1+2g}$.}
    \label{Fig_Ratio_Int_FiniteT}
\end{figure}

$\delta_2(g)$ can be related to the compressibility $\kappa=\frac{1}{V}\frac{\partial N}{\partial \mu}$ of the Fermi liquid as follows. If we extend the region $A$ to the whole system, we have in the grand canonical ensemble
\begin{align}
    \jkh{N^2}_c=\frac{1}{\beta}\frac{\partial N}{\partial \mu}=\frac{\pi R^2}{\beta}\kappa.
\end{align}
The compressibility of non-interacting fermions is the density of states at the Fermi surface:
\begin{align}
    \kappa_0=N(0)=\frac{k_F}{2\pi v_F}.
\end{align}
For interacting fermion systems, the compressibility can be computed within the framework of Landau’s Fermi liquid theory. Given the isotropic interaction described by Eq.~\eqref{MB_Eq_Interaction}, the only non-zero dimensionless Landau parameter is $F_0$. Consequently, the compressibility is given by~\cite{coleman2015introduction}
\begin{align}
    \kappa=\frac{N(0)}{1+F_0}.
\end{align}
Since there is no mass renormalization, we still have $N(0)$ in the expression. It follows that
\begin{align}
    \delta_2(g)=\frac{1}{1+F_0}=\frac{1}{1+2g},
\end{align}
and
\begin{align}
    Z(\theta,\beta)\simeq\exp\kh{-\theta^2 \frac{k_F}{4\beta v_F} \frac{1}{1+2g}R^2}.
    \label{Z_finite_T}
\end{align}

We also calculate $\delta_2(g)$ via numerical diagonalization, with the results shown in Fig.~\ref{Fig_Ratio_Int_FiniteT}. The data exhibit an excellent fit to $\frac{1}{1+2g}$ across the entire range, reaffirming the relation $F_0 = 2g$.

\section{Disorder operators in non-Fermi liquids}\label{Sec_NFL}
In this section, we study the disorder operator in simple models of non-Fermi liquids, again using the bosonization formalism.

\subsection{Fermions coupled to a critical scalar} \label{Sec_NFL_1}
First, we consider a non-Fermi liquid model introduced in Ref.~\cite{delacretaz2022nonlinear} by adding a bosonic order parameter $\Phi(t,\vb{x})$ to the non-interacting fermions. For this problem, it is more convenient to use the Lagrangian formalism, although we go back to the Hamiltonian formalism for numerical diagonalizations.

Following Ref.~\cite{delacretaz2022nonlinear}, the free theory after bosonization is described by the following Lagrangian: 
\begin{align}
    \mathcal{L}_0=-\frac{k_F}{8\pi^2} \int \diff{\theta}\, \vu{n}_{\theta}\cdot \nabla \phi \kh{\dot{\phi}+v_F \vu{n}_{\theta}\cdot\nabla\phi},
\end{align}
where $\theta$ is a variable on the Fermi surface and $\vu{n}_{\theta}$ is the unit vector in the direction of $\theta$. The fermion density is related to the field $\phi$ by
\begin{align}
    \rho(t,\vb{x})=\int\diff\theta\, \rho(t,\vb{x},\theta)=\frac{k_F}{(2\pi)^2} \int \diff\theta\, \vu{n}_{\theta}\cdot \nabla \phi.
\end{align}
$\rho(t,\vb{x},\theta)$ is the fermion density corresponding to the Fermi momentum $k_F \vu{n}_{\theta}$.

The Lagrangian density of the boson field is
\begin{align}
    \mathcal{L}_B=\frac{1}{2v_{\Phi}^2} (\partial_t \Phi)^2-\frac{1}{2} (\nabla \Phi)^2-\frac{1}{2} k_{\Phi}^2 \Phi^2,\label{NFL_Eq_LB}
\end{align}
where $v_\Phi$ is the boson velocity and $k_\Phi^2$ is the bare mass. $\Phi(t,\vb{x})$ couples to the fermion density by
\begin{align}
    \mathcal{L}_{\textrm{int}}=\lambda \Phi(t,\vb{x}) \rho\kh{t,\vb{x}}.\label{NFL_Eq_Lint}
\end{align}
The full low-energy effective Lagrangian reads
\begin{align}
    \mathcal{L}=\mathcal{L}_0+\mathcal{L}_B+\mathcal{L}_{\textrm{int}}.
    \label{Eq_Lagrangian_NFL_1}
\end{align}

Since the coupling $\mathcal{L}_{\rm int}$ is linear in $\rho$, the full theory is still quadratic in fermion density and can be solved exactly within bosonization. When doing numerical diagonalization, we will set $v_{\Phi}=v_F$ for simplicity, since $v_{\Phi}$ does not affect the behavior of $G(q)$ near $q=0$. In fact, the $(\partial_t \Phi)^2$ term is irrelevant in this setting.

First, we discuss the phase diagram as a function of the boson mass $k_\Phi^2$. The fermions can be integrated to give an effective action for the bosonic field, with an effective mass $k_\Phi^2-k_c^2$. Here we define $k_c^2=\frac{k_F}{2\pi v_F} \lambda^2$. The boson is gapped if $k_{\Phi}^2>k_c^2$, and when $k_{\Phi}^2<k_c^2$, the boson has a negative mass, indicating an instability of boson condensation. Physically, when the boson condenses (so $\Phi$ has a finite expectation value), the coupling causes a change in the fermion density. In other words, the system enters an inhomogeneous ``phase separation" state. 

The critical point~\footnote{We note that Ref.~\cite{delacretaz2022nonlinear} presents this relation with an opposite sign, which would cause an instability in the theory even at the critical point.} is located at
\begin{equation}
        k_{\Phi}^2=k_c^2 =\frac{k_F}{2\pi v_F} \lambda^2,\label{NFL_Eq_Gapless}
\end{equation}
where both fermions and the boson are gapless. As shown in Ref.~\cite{delacretaz2022nonlinear}, at the critical point, the boson is Landau damped with dynamical exponent $z=3$. In Appendix~\ref{App_Spec_NFL}, we systematically study the spectrum of the theory using field-theoretic bosonization.

\begin{figure}[ht!]
    \centering
    \includegraphics[width=\columnwidth]{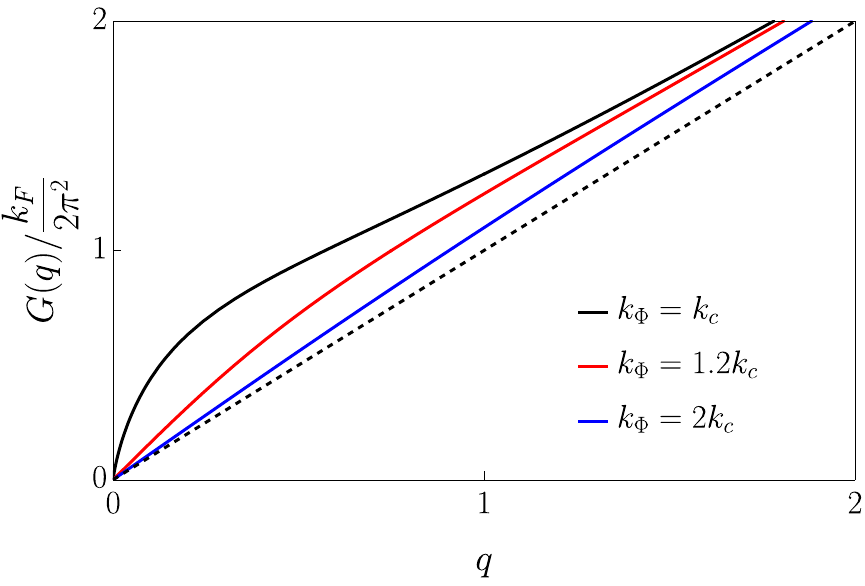}
    \caption{$G(q)/\frac{k_F}{2\pi^2}$ vs $q$ for $k_c=1$ and three different values of $k_{\Phi}$. The critical case is depicted by the solid black line, while the free fermion case is shown with a dashed line for comparison.}
    \label{Fig_NFL1_Gq}
\end{figure}

We now study the structure factor $G(q)$. We will focus on the small $q$ behavior, which is responsible for the scaling form of the disorder operator. In addition, when $q$ is large, the coupling to boson field becomes unimportant, and $G(q)$ approaches the free fermion expression. For $k_{\Phi}^2>k_c^2$, finite-patch numerical diagonalization shows that $G(q)$ is still linear in $q$ with a modified slope at small momentum $q$, which is expected for an interacting Fermi liquid. For a fixed $k_c^2$, the slope (and thus the coefficient of the leading $R\ln R$ term) increases as $k_{\Phi}^2$ decreases toward the critical point, as shown in Fig.~\ref{Fig_NFL1_Gq}. However, at the critical point, $G(q\rightarrow 0)$ exhibits a logarithmic dependence on the number of patches $N$, signaling a qualitative change of the scaling.

To better understand the $q\rightarrow 0$ behavior, we consider the $v_\Phi\rightarrow \infty$ limit, where the bosons effectively mediate a static interaction between the fermions. One can then directly apply Eq.~\eqref{BC_Eq_G} to find
\begin{align}
    G(q)=\frac{k_F q}{2\pi^2}\delta_1\kh{-\frac12 \frac{k_c^2}{q^2+k_{\Phi}^2}}.\label{NFL_Eq_Gq}
\end{align}
As $k_\Phi \rightarrow k_c^+$, the effective $g$ approaches $-1/2$, and $F_0\rightarrow -1$, signaling the divergence in compressibility and a phase separation instability.

\begin{figure}[ht]
    \centering
    \includegraphics[width=\columnwidth]{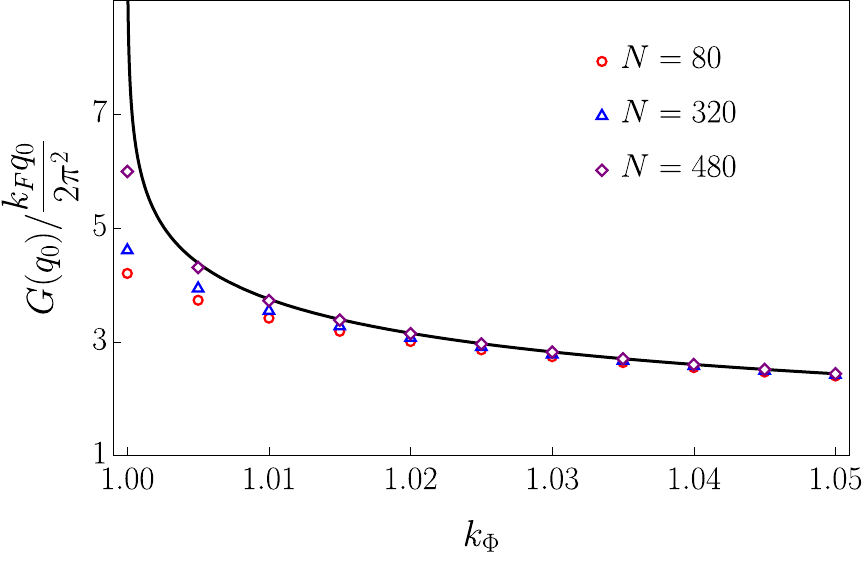}
    \caption{$G(q_0)/\frac{k_F q_0}{2\pi^2}$ for $k_c=1$ and $q_0=10^{-3}$. We show numerical results of $80,320$, and $480$ patches. The solid black line is the field-theoretic result Eq.~\eqref{NFL_Eq_Gq}.}
    \label{Fig_NFL1_Slope}
\end{figure}

Near $g=-\frac12$, we have
\begin{align}
    \delta_1(g)\simeq -1-\ln\kh{g+\frac{1}{2}}.
\end{align}
Therefore, for $k_\Phi^2>k_c^2$, $G(q)$ is still linear in $q$ when $q\rightarrow 0$:
\begin{align}
    G(q)\simeq\frac{k_F q}{2\pi^2} \delta_1\kh{-\frac{k_c^2}{2k_{\Phi}^2}},
\end{align}
in perfect agreement with numerical results. However, the slope diverges logarithmically as $k_\Phi^2\rightarrow k_c^2$:  
\begin{equation}
    \frac{G(q)}{q}\simeq -\frac{k_F}{2\pi^2}\ln\kh{1- \frac{k_c^2}{k_\Phi^2}}.
    \label{Eq_RlnR_coef}
\end{equation}
We compare the field theory results with those obtained from finite patch numerical diagonalizations in Fig.~\ref{Fig_NFL1_Slope}, for a fixed small $q_0$. Slightly away from the critical point, the finite-patch results converge well to the value given by Eq.~\eqref{NFL_Eq_Gq}. As the system approaches the critical point, the convergence of the finite-patch calculations becomes significantly slower.

At the critical point $k_\Phi^2=k_c^2$, we find
\begin{align}
    G(q)=-\frac{k_F}{\pi^2} q \ln q + \bO(q).
\end{align}
$G(q)$ is logarithmically enhanced as $q\rightarrow 0$, although one still has $G(q\rightarrow 0)=0$. We therefore expect an enhancement of the charge fluctuations, which makes sense as the system is at the onset of an instability towards phase separation. Performing the integral in Eq.~\eqref{DO_Eq_NAbyInt} (see Appendix~\ref{integral} for details), the leading-order term of $\jkh{N_A^2}_c$ is
\begin{align}
    \jkh{N_A^2}_c\simeq \frac{k_F}{\pi^2}R\ln^2\frac{R}{\alpha}.
\end{align}
This is one of the main results of the work, that the charge fluctuations exhibit $R\ln^2 R$ scaling in the NFL state.

We can easily generalize the results to the problem of spin-$1/2$ fermions coupled to ferromagnetic critical fluctuations. First, the free theory can be bosonized directly with two bosonic fields $\phi_\uparrow$ and $\phi_\downarrow$, corresponding to spin $\uparrow$ and $\downarrow$ fermions, respectively. Since the bosonized free theory is quadratic, we define the charge and the spin modes:
\begin{equation}
    \phi_c=\frac{1}{\sqrt{2}}(\phi_\uparrow + \phi_\downarrow), \phi_s=\frac{1}{\sqrt{2}}(\phi_\uparrow - \phi_\downarrow).
\end{equation}
The free part of the Lagrangian for the fermions becomes
\begin{align}
    \mathcal{L}_0=-\frac{k_F}{8\pi^2}\sum_{a=c,s} \int \diff\theta\, \vu{n}_{\theta}\cdot \nabla \phi_a \kh{\dot{\phi}_a+v_F \vu{n}_{\theta}\cdot\nabla\phi_a}.
\end{align}
The charge and spin densities are
\begin{align}
    \rho&=\frac{\sqrt{2}k_F}{(2\pi)^2}\int \diff\theta\, \vu{n}_{\theta}\cdot \nabla \phi_c(\vb{x}),\\
    \rho_s&=\frac12\frac{\sqrt{2}k_F}{(2\pi)^2}\int \diff\theta\, \vu{n}_{\theta}\cdot \nabla \phi_s(\vb{x}).
\end{align}

Ferromagnetic spin fluctuations can be described by a real scalar field $\Phi$ as before, the action of which still takes the form Eq.~\eqref{NFL_Eq_LB}. The coupling term is modified to 
\begin{equation}
    \mathcal{L}_{\textrm{int}}=\lambda \Phi(\rho_\uparrow - \rho_\downarrow) =\frac{\sqrt{2} \lambda k_F}{(2\pi)^2}\Phi\int \diff\theta\, \vu{n}_{\theta}\cdot \nabla \phi_s.
\end{equation}

Within bosonization, the charge sector remains free, and we have
\begin{align}
    \langle N_A^2\rangle_c &\simeq \frac{2k_F}{\pi^2}R\ln R. \label{FMNFL_N}
\end{align}
On the other hand, it is clear that the spin sector is exactly mapped to the theory Eq.~\eqref{Eq_Lagrangian_NFL_1}. We thus conclude that at the critical point
\begin{align}
    \langle (S^z_A)^2\rangle_c &\simeq  \frac{k_F}{2\pi^2}R\ln^2 R.
\end{align}
In the paramagnetic phase $k_\Phi^2>\frac{k_F}{\pi v_F}\lambda^2$, we have $\jkh{(S^z_A)^2}_c\propto R\ln R$ with a coefficient given essentially by Eq.~\eqref{Eq_RlnR_coef} (with appropriate redefinition of parameters).

\subsection{Composite Fermi liquid}
Another non-Fermi liquid model that can be treated effectively using bosonization is the Halperin-Lee-Read (HLR) theory~\cite{HLR} of composite Fermil liquid in the half-filled Landau level~\cite{kwon1994gauge}. In this theory, the (charge-neutral) composite fermions are coupled to an emergent U(1) gauge field $a_{\mu}=(a_0,\vb{a})$. Importantly, the physical fermion density is represented by the magnetic flux: 
\begin{align}
    \rho_{\textrm{phy}}(\vb{r})=\frac{1}{4\pi} \nabla\times\vb{a}=\frac{1}{4\pi}(\partial_1 a_2-\partial_2 a_1).
\end{align}

The low-energy effective Lagrangian reads
\begin{align}
    \mathcal{L}=\mathcal{L}_0+\mathcal{L}_G+\mathcal{L}_{\textrm{int}},
\end{align}
where $\mathcal{L}_0$ is the Lagrangian of bosonized composite fermions and $\mathcal{L}_G$ is the Chern-Simons term. In the Coulomb gauge $\nabla\cdot\vb{a}=0$, we have
\begin{align}
    \mathcal{L}_G=\frac{1}{8\pi} a\diff a=\frac{1}{4\pi} a_0(\nabla\times\vb{a}).
\end{align}
$\mathcal{L}_{\textrm{int}}$ contains two terms. The first resembles a generalization to Eq.~\eqref{NFL_Eq_Lint} with gauge field coupled to the charge currents,
\begin{align}
    \int\frac{\Diff{2}{\vb{q}}\diff\omega}{(2\pi)^3}\diff\theta\, \rho(\omega,\vb{q},\theta)(a_0(-\omega,-\vb{q})+v_F \vu{n}_{\theta}\cdot \vb{a}(-\omega,-\vb{q})),
\end{align}
while the second term is generated by integrating out fields deep inside the Fermi sea:
\begin{align}
    -\int\frac{\Diff{2}{\vb{q}}\diff\omega}{(2\pi)^3}\frac{k_F v_F}{8\pi}\vb{a}(\omega,\vb{q})\cdot\vb{a}(-\omega,-\vb{q}).
\end{align}

The structure factor can be obtained from the gauge propagator after integrating out the composite fermions
\begin{align}
    \ii G(\omega,q)=\frac{k_F}{2\pi v_F} \frac{q^2 f_1(s)}{q^2-4k_F^2 f_1(s)f_2(s)},\label{NFL2_Eq_Full}
\end{align}
where $s=\frac{\omega}{v_F q}$. $f_1(s)$ and $f_2(s)$ are the longitudinal and transverse susceptibility functions respectively:
\begin{align}
    f_1(s)=1-\frac{\abs{s}}{\sqrt{1-s^2}},
\end{align}
and
\begin{align}
    f_2(s)=-s^2+\abs{s}\sqrt{s^2-1}.
\end{align}
For $\omega\ll v_Fq$, we obtain
\begin{align}
    \ii G(\omega,q)=\frac{k_F}{2\pi v_F} \frac{1}{1-4\ii k_F^2 \frac{\omega}{v_F q^3}},\label{NFL2_Eq_Small}
\end{align}
in agreement with RPA results of the HLR theory.

To obtain the equal-time correlation, we need to integrate over $\omega$. However, direct integration of the full expression in Eq.~\eqref{NFL2_Eq_Full} leads to divergences that are difficult to control. Instead, we begin with Eq.~\eqref{NFL2_Eq_Small}. For $q$ small, it yields $G(q)\propto q^3 \ln (1/ql_B)$, as noted in Ref.~\cite{ReadCFL}. Performing the integral in Eq.~\eqref{DO_Eq_NAbyInt} (see Appendix~\ref{integral} for details), one finds that $\jkh{N_A^2}_c$ is proportional to $R$ in the leading order.

\section{Discussions}\label{Sec_discussions}
We now compare our analytical theory with the numerical study Ref.~\cite{FermionDisorderOp} of a lattice model with two layers of spin-1/2 fermions coupled to Ising spins governed by a transverse-field ferromagnetic Ising Hamiltonian, which uses large-scale sign-free Monte Carlo simulations. First we briefly review the numerical results. When the transverse field $h$ is greater than a critical value $h_c$, the model is in the paramagnetic phase, with a Fermi liquid ground state (the Ising spins are gapped). For $h<h_c$, the Ising spins enter the ferromagnetic state and the spin-up and -down Fermi surfaces split due to the magnetization. At $h=h_c$, the fermions are coupled to the critical spin fluctuations realizing a NFL state~\cite{NFLFM, Xu_2020}. While not exactly identical, the low-energy theory of this system is very similar to the one studied in Sec.~\ref{Sec_NFL_1}, just with more flavors of fermions.

The model is invariant under both $\U_{\textrm{charge}}$ and $\U_{\textrm{spin}}$, and one can measure both charge $Q$ and spin $S^z$ disorder operators, which will be denoted by $Z_c$ and $Z_s$ respectively below. Let us discuss the main numerical findings in light of the theory of disorder operators in the NFL state developed in Sec.~\ref{Sec_NFL}.
\begin{enumerate}
    \item Within available system sizes, the disorder operator $\ln \abs{Z_{c/s}(\theta)}$ is proportional to $\theta^2$ for all $\theta\in [0,\pi]$. This crucial observation suggests that the charge/spin fluctuations obey Gaussian distribution to a good approximation, supporting the fundamental assumption of the (linear) bosonization approach used in the theoretical calculations.
    \item The charge disorder operator $Z_c$ shows little change as $h$ approaches the critical value from above (i.e. the paramagnetic side), including at the critical point. This is in reasonable agreement with the theory where the charge disorder parameter does not change with $k_\Phi^2$ as long as $k_\Phi^2\geq k_c^2$ (see Eq.~\eqref{FMNFL_N}). On the other hand, $-\ln \abs{Z_s}$ shows a sharp peak at $h=h_c$, which is also consistent with the theoretical results shown in Fig.~\ref{Fig_NFL1_Slope}.
    \item $-\ln \abs{Z_s}$ at $h=h_c$ shows clear deviation from the non-interacting values. The deviation was observed to grow with the size of the region with no sign of saturation up to the largest available size. Similar deviation was seen in the second R\'enyi entropy. On the other hand, $-\ln \abs{Z_c}$ remains close to the non-interacting values. Our theory indeed shows that $\langle (S^z_A)^2\rangle_c \propto l_A\ln^2 l_A$ at the critical point, thus giving a natural explanation of the diverging deviation away from the free case (which should scale as $l_A\ln l_A$).
\end{enumerate}

To summarize, all the qualitative features of the numerical observations about charge and spin disorder operators in Ref.~\cite{FermionDisorderOp} are captured by the theoretical results obtained in Sec.~\ref{Sec_NFL}, lending strong support to the validity of the bosonization approach.

Let us now discuss future directions.

One important question is to access the nonlinear effects, e.g. coming from the curvature of the Fermi surface. It is well known that the nonlinear effects are important for certain aspects of the system, for example they are expected to cause the boson dynamical exponent $z$ to deviate from $z=3$. It would be worthwhile to investigate the nonlinear corrections more systematically, by e.g. computing $\jkh{N_A^3}_c$ using a different method, such as the nonlinear completion of bosonization recently proposed in Ref.~\cite{delacretaz2022nonlinear}, or direct diagrammatic calculations in the fermionic formulation.

The multidimensional bosonization method was used to compute EE in a Fermi liquid in Ref.~\cite{ding2012entanglement}, and it was shown that the EE has the same $l_A\ln l_A$ scaling as the one in a Fermi gas, with the coefficient given by the Widom formula. It will be interesting to extend the bosonization method to computing entanglement entropy for NFL models~\cite{EECSL, EECFL1, EECFL2, swingle2013universal}. In particular, the fact that $\jkh{N_A^2}$ shows $l_A\ln^2 l_A$ dependence in the NFL model in Sec.~\ref{Sec_NFL_1} suggests that the entanglement entropy might also exhibit a different scaling. This possibility was also suggested by the numerical results in Ref.~\cite{FermionDisorderOp}, where it is found that the second R\'enyi entropy $S^{(2)}_A$ shows deviations from the non-interacting result that grow with $l_A$.

It will also be interesting to consider other models of NFLs, for example spin-$1/2$ fermions coupled to a U(1) gauge field, which describes the low-energy physics of certain gapless spin liquids with the spinon Fermi surface, or the Fermi surface coupled to anti-ferromagnetic spin fluctuations. 

\emph{Note added:} During the finalization of this work, we became aware of an upcoming related work \cite{Wu2024} to appear in the same arXiv listing.

\begin{acknowledgments}  
We thank Zhen Bi, Dominic Else, Ruihua Fan, Yingfei Gu, Max Metlitski, Dam Thanh Son, and Xiao-Chuan Wu for illuminating conversations. M.C. is especially grateful to Max Metlitski for a discussion on the composite Fermi liquid problem, and Xiao-Chuan Wu for communicating unpublished results and pointing out an error in our draft. M.C. acknowledges support from NSF Grant No. DMR-1846109.

\end{acknowledgments}

\bibliography{ref.bib}

\onecolumngrid

\appendix

\section{Evaluation of integrals}\label{integral}
In this section, we address the leading $R$ dependence of $\jkh{N_A^2}_c$ in Eq.~\eqref{DO_Eq_NAbyInt1}:
\begin{align}
    \jkh{N_A^2}_c=2\pi R \int_{0}^{\frac{R}{\alpha}} \frac{G(x/R)}{x/R}J_1^2(x)\diff x,
\end{align}
where $J_1$ is the Bessel function of the first kind, and $\frac{1}{\alpha}$ is a high-$q$ cut-off. For $R\rightarrow\infty$, the leading $R$ dependence is determined by the behavior of $G(q)$ as $q\rightarrow 0$. Below we treat two cases separately: the first one is when $G(q)\sim q^n$ for $n\geq 0$. The second case is when $G(q)\sim q^n \ln q$.

\subsection{Power law}
Supposing $G(q)$ takes the power-law form $G(q)\simeq a q^n (n\geq 0)$ as $q\rightarrow 0$, we have
\begin{align}
    \jkh{N_A^2}_c\simeq 2a\pi R^{2-n} \int_{0}^{\frac{R}{\alpha}} x^{n-1} J_1^2(x)\diff x.
\end{align}
For $0\leq n< 1$, the integral converges as we take the upper limit $\frac{R}{\alpha}$ to infinity, and
\begin{align}
    \jkh{N_A^2}_c\simeq 2a\pi R^{2-n} \int_{0}^{\infty} x^{n-1} J_1^2(x)\diff x=a\sqrt{\pi} \frac{\Gamma\kh{\frac{1}{2}-\frac{n}{2}}\Gamma\kh{1+\frac{n}{2}}}{\Gamma\kh{1-\frac{n}{2}}\Gamma\kh{2-\frac{n}{2}}} R^{2-n}.
\end{align}
Thus, we get volume-law scaling $\jkh{N_A^2}_c\simeq a\pi R^2$ if we take $n=0$.

For $n\geq 1$, the integral diverges, and the leading contribution is then related to the behavior of the integrand for $x$ large. Using the asymptotic form of Bessel function $J_1^2(x)\simeq \frac{1}{\pi x}(1-\sin 2x)$ as $x\rightarrow\infty$, for $n>1$, we have
\begin{align}
    \jkh{N_A^2}_c&\simeq 2a\pi R^{2-n} \int_{x_0}^{\frac{R}{\alpha}} x^{n-1} J_1^2(x)\diff x\\
    &\simeq 2a R^{2-n} \int_{x_0}^{\frac{R}{\alpha}} x^{n-2}(1-\sin 2x) \diff x\\
    &\simeq 2a R^{2-n} \frac{1}{n-1}\kh{\frac{R}{\alpha}}^{n-1}\\
    &\simeq \frac{2a}{n-1} \alpha^{1-n} R.\label{Scaling_Eq_PowerlawN1}
\end{align}
where $x_0$ is a sufficiently large number as a lower cut off and we neglect the oscillatory $\sin 2x$ term in the third line.

The $n=1$ case is different and needs a separate calculation:
\begin{align}
    \jkh{N_A^2}_c&\simeq 2a R \int_{x_0}^{\frac{R}{\alpha}} \frac{1-\sin 2x}{x} \diff x \simeq 2aR\ln\frac{R}{\alpha}.
\end{align}

\subsection{Power law with logarithmic enhancement}
The more interesting situation is when $G(q)$ is enhanced by a logarithmic factor $G(q)\sim a q^n \ln q$ near $q=0$.
For $0\leq n <1$, the integral with an extra $\ln x$ still converges:
\begin{align}
    I_n=\int_0^{\infty} x^{n-1}\ln x J_1^2(x)\diff x<\infty.
\end{align}\
Then, we have
\begin{align}
    \jkh{N_A^2}_c&\simeq 2a\pi R^{2-n} \int_{0}^{\frac{R}{\alpha}} x^{n-1} \ln\frac{x}{R} J_1^2(x)\diff x\\
    &=2a\pi R^{2-n} \int_{0}^{\infty} x^{n-1} \ln x J_1^2(x)\diff x-2a\pi R^{2-n}\ln R \int_{0}^{\infty} x^{n-1} J_1^2(x)\diff x\\
    &\simeq -a\sqrt{\pi} \frac{\Gamma\kh{\frac{1}{2}-\frac{n}{2}}\Gamma\kh{1+\frac{n}{2}}}{\Gamma\kh{1-\frac{n}{2}}\Gamma\kh{2-\frac{n}{2}}} R^{2-n}\ln R+2\pi I_n R^{2-n}.
\end{align}
We find that the leading-order scaling also gets enhanced by a logarithmic factor in this case.

For $n>1$, repeating the steps in Eq.~\eqref{Scaling_Eq_PowerlawN1}, we find
\begin{align}
    \jkh{N_A^2}_c&\simeq 2a\pi R^{2-n} \int_{0}^{\frac{R}{\alpha}} x^{n-1} \ln\kh{\frac{x}{R}} J_1^2(x)\diff x\\
    &= 2a\pi R^{2-n} \int_{0}^{\frac{R}{\alpha}} x^{n-1} \ln x J_1^2(x)\diff x-
    2a\pi R^{2-n}\ln R \int_{0}^{\frac{R}{\alpha}} x^{n-1} J_1^2(x)\diff x\\
    &\simeq 2a R^{2-n} \int_{x_0}^{\frac{R}{\alpha}} x^{n-2} \ln x \diff x-
    2a R^{2-n}\ln R \int_{x_0}^{\frac{R}{\alpha}} x^{n-2} \diff x\\
    &\simeq \frac{2aR^{2-n}}{n-1} \fkh{\kh{\ln\frac{R}{\alpha}-\frac{1}{n-1}}\kh{\frac{R}{\alpha}}^{n-1} - \ln R\kh{\frac{R}{\alpha}}^{n-1}}\\
    &\simeq \frac{2a}{n-1} \alpha^{1-n} \kh{\ln\frac{1}{\alpha}-\frac{1}{n-1}} R.
\end{align}
Unlike in the $0\leq n<1$ case, the coefficient of the $R\ln R$ term vanishes, and the actual leading-order term is linear in $R$. In other words, the logarithmic factor in $G(q)$ does not change the scaling of $\jkh{N_A^2}_c$ for $n>1$.

The $n=1$ case is again different. We have
\begin{align}
    \jkh{N_A^2}_c&\simeq 2a R \int_{x_0}^{\frac{R}{\alpha}} \frac{\ln x}{x} \diff x-
    2a R\ln R \int_{x_0}^{\frac{R}{\alpha}} \frac{1}{x} \diff x\\
    &\simeq a R\ln^2\frac{R}{\alpha}-2aR\ln R\ln\frac{R}{\alpha}\\
    &\simeq -a R\ln^2\frac{R}{\alpha}.
\end{align}
We find an extra logarithmic factor and the scaling becomes $R\ln^2 R$.

\section{Explicit diagonalization of \texorpdfstring{$\mathcal{H}(q)$}{H(q)}}\label{App_Diag}
To diagonalize the Hamiltonian $\mathcal{H}(q)$ in Eq.~\eqref{MB_Eq_Hq}, we first exploit the $\mathbb{Z}_2$ symmetry $i\leftrightarrow -i$. For $i>0$, we make the following change of basis:
\begin{align}
    \begin{aligned}
        b_{i,\pm}&=\frac{1}{\sqrt{2}}(b_i \pm b_{-i}),\\
        a_{i,\pm}&=\frac{1}{\sqrt{2}}(a_i \pm a_{-i}).
    \end{aligned}
\end{align}
There is no mixing between the $+$ modes and the $-$ modes, so $\mathcal{H}(q)=H_+ + H_-$. In addition, the $-$ modes are already diagonalized
\begin{align}
    H_-=\sum_{i>0} s_i \kh{b^{\dagger}_{i,-}b_{i,-}+a^{\dagger}_{i,-}a_{i,-}}.
\end{align}
Since only the $+$ modes appear in the fermion density, we can focus on the $H_+$ part:
\begin{align}
    H_+ = &\sum_{i>0} s_i \kh{b^{\dagger}_{i}b_{i}+a^{\dagger}_{i}a_{i}}+\kh{b^{\dagger}_0 b_0+a^{\dagger}_0 a_0}
    +4\tilde{g}\sum_{i,j\geq 0} \sqrt{s_i s_j} \kh{b^{\dagger}_{i}b_{j}+a^{\dagger}_{i}a_{j}+a_{i}b_{j}+b^{\dagger}_{i} a^{\dagger}_{j}}.
\end{align}
For simplicity, we have suppressed the $+$ label on the modes here and for the rest of the section. $s_0$ is redefined to be $s_0=\frac{1}{2}$ to account for the absence of $\sqrt{2}$ coefficients in $b_0$ and $a_0$ modes.

$H_+$ can be further diagonalized with a Bogoliubov transformation of the following form:
\begin{align}
    \begin{aligned}
        b_i&=\sum_{l\geq 0}\kh{M_{il}\beta_l+N_{il}\alpha^{\dagger}_l}\\
        a_i&=\sum_{l\geq 0}\kh{M_{il}\alpha_l+N_{il}\beta^{\dagger}_l} \label{Diag_Eq_Bogoliubov}
    \end{aligned},
\end{align}
where $M$ and $N$ are real matrices. The canonical commutation relations enforce that
\begin{align}
    \begin{aligned}
        \sum_{l\geq 0}(M_{il}M_{jl}-N_{il}N_{jl}) &=\delta_{i,j}\\
        \sum_{l\geq 0}(M_{il}N_{jl}-N_{il}M_{jl}) &=0\label{Diag_Eq_Commutator}
    \end{aligned}.
\end{align}
The Hamiltonian after the transformation is assumed to be in the diagonal form:
\begin{align}
    H_+=\sum_{l\geq 0}S_l\kh{\beta^{\dagger}_l \beta_l+\alpha^{\dagger}_l\alpha_l}+E_0.
\end{align}
To find the equation for the Bogoliubov coefficients, consider the commutator
\begin{align}
    \left[ b_i,H_+ \right] = \sum_{l\geq 0} S_l\kh{M_{il}\beta_l-N_{il}\alpha^{\dagger}_l}
    =\begin{cases}
        s_i b_i + 4\tilde{g}\sum_{j\geq 0}\sqrt{s_i s_j}\kh{b_j+a^{\dagger}_j} & i>0\\
        b_0 + 4\tilde{g}\sum_{j\geq 0}\sqrt{s_0 s_j}\kh{b_j+a^{\dagger}_j} & i=0
    \end{cases}.
\end{align}
Using the definition of the Bogoliubov transformation Eq.~\eqref{Diag_Eq_Bogoliubov} and comparing the coefficients of each mode, we have the following set of equations, for $i>0$:
\begin{align}
    \begin{aligned}
        (S_l-s_i) M_{il}&=4\sqrt{s_i} \tilde{g} C_l\\
        (S_l+s_i) N_{il}&=-4\sqrt{s_i} \tilde{g} C_l
    \end{aligned},\label{Diag_Eq_Mpositive}
\end{align}
and for $i=0$
\begin{align}
    \begin{aligned}
        (S_l-1) M_{0l}&=2\sqrt{2} \tilde{g} C_l\\
        (S_l+1) N_{0l}&=-2\sqrt{2} \tilde{g} C_l
    \end{aligned},\label{Diag_Eq_M0}
\end{align}
where $C_l$ is defined to be
\begin{align}
    C_l=\sum_{j\geq 0} \sqrt{s_j}(M_{jl}+N_{jl}) \label{Diag_Eq_Cl}
\end{align}

If we assume that $S_l-s_i$ are all non-zero, then we can safely divide both sides of Eq.~\eqref{Diag_Eq_Mpositive} by $S_l\pm s_i$. Then, the solutions of $M_{il}$ and $N_{il}$ for $i>0$ are
\begin{align}
    \begin{aligned}
        M_{il}&=\tilde{g}\frac{4\sqrt{s_i}C_l}{S_l-s_i}\\
        N_{il}&=-\tilde{g}\frac{4\sqrt{s_i}C_l}{S_l+s_i}
    \end{aligned}.\label{Diag_Eq_MpositiveSol}
\end{align}
Similarly, for $i=0$, we need to assume $S_l\neq 1$, and
\begin{align}
    \begin{aligned}
        M_{0l}&=\tilde{g}\frac{2\sqrt{2}C_l}{S_l-1}\\
        N_{0l}&=-\tilde{g}\frac{2\sqrt{2}C_l}{S_l+1}
    \end{aligned}.\label{Diag_Eq_M0Sol}
\end{align}

The eigenenergies $S_l$ can be determined from the consistency equation, which is obtained by substituting the solutions into the definition of $C_l$ [Eq.~\eqref{Diag_Eq_Cl}]:
\begin{align}
    8\tilde{g}\sum_{j>0} \frac{s_j^2}{S_l^2-s_j^2} + 4\tilde{g}\frac{1}{S_l^2-1}=1.\label{Diag_Eq_Consistency}
\end{align}
This equation does have $\frac{N}{4}$ solutions not equal to $s_i$ and $1$, which justifies the assumption. Actually, for $N$ large, there are two kinds of solutions. The first kind is related to the particle-hole continuum with $S_l=s_l+\delta_l<1$, where $\delta_l$ is a correction of order $\bO(N^{-1})$. After taking the $N\rightarrow\infty$ limit, $S_l$ just converge to the bare values $s_l$.
The second kind is a single collective mode solution $S_0>1$. We can find this collective mode by converting the summation in Eq.~\eqref{Diag_Eq_Consistency} into an integral: 
\begin{align}
    \frac{4g}{\pi}\int_0^{\frac{\pi}{2}} \frac{\cos^2 x}{S_0^2-\cos^2 x}\diff x=1,\label{Diag_Eq_S0Int}
\end{align}
and the solution is given by
\begin{align}
    S_0=\frac{1+2g}{\sqrt{1+4g}}.
\end{align}

Starting from this point, a more efficient numerical diagonalization method is to first solve the consistency equation Eq.~\eqref{Diag_Eq_Consistency} for all the eigenenergies, and then substitute Eq.~\eqref{Diag_Eq_MpositiveSol} and Eq.~\eqref{Diag_Eq_M0Sol} into the restriction Eq.~\eqref{Diag_Eq_Commutator} to get a set of linear equations in $C_l$ that determines all the coefficients.

\section{Spectrum of critical boson non-Fermi liquid}\label{App_Spec_NFL}
In this section, we study the spectrum of the critical boson non-Fermi liquid theory in Sec.~\ref{Sec_NFL_1} analytically using field-theoretic bosonization~\cite{delacretaz2022nonlinear}. The full Lagrangian is 
\begin{align}
    \mathcal{L}=-\frac{k_F}{8\pi^2} \int \diff\theta\, \vu{n}_{\theta}\cdot \nabla \phi \kh{\dot{\phi}+v_F \vu{n}_{\theta}\cdot\nabla\phi}+\frac{1}{2v_{\Phi}^2} (\partial_t \Phi)^2-\frac{1}{2} (\nabla \Phi)^2-\frac{1}{2} k_{\Phi}^2 \Phi^2+\lambda\rho(t,\vb{x})\Phi(t,\vb{x}),
\end{align}

Integrating out $\Phi$ gives an effective density-density interaction of the fermions:
\begin{align}
    -\frac{1}{2}\lambda^2\int \diff{t}\Diff{2}{\vb{x}}\, \rho \frac{1}{-\frac{\partial_t^2}{v_{\Phi}^2}+\nabla^2-k_{\Phi}^2} \rho.
\end{align}
After Fourier transformation on $t$ and $\vb{x}$, we have
\begin{align}
    S_{\textrm{eff}}=\frac{1}{2}\int\frac{\diff{\omega}}{2\pi}\frac{\Diff{2}{\vb{q}}}{(2\pi)^2}
    \diff{\theta_1}\diff{\theta_2}\phi(\omega,\vb{q},\theta_1) K \phi(-\omega,-\vb{q},\theta_2),\label{NFL_Eq_Seff}
\end{align}
where the kernel $K$ is
\begin{align}
    K=&\frac{k_F}{4\pi^2} (\vu{n}_{\theta_1}\cdot \vb{q}) (\omega-v_F \vu{n}_{\theta_2}\cdot \vb{q}) \delta(\theta_1-\theta_2)
    -\lambda^2 \frac{k_F^2}{(2\pi)^4} (\vu{n}_{\theta_1}\cdot \vb{q}) (\vu{n}_{\theta_2}\cdot \vb{q}) \frac{1}{\frac{\omega^2}{v_{\Phi}^2}-q^2-k_{\Phi}^2}.
\end{align}

To diagonalize the effective action in the $\theta$ coordinate, we expand $\phi(\omega, \vb{q}, \theta)$ in a Fourier series of $\theta$, with coefficients $\phi(\omega, \vb{q}, l)$ for $l\in \mathbb{Z}$. The action becomes
\begin{align}
    S_{\textrm{eff}}=&\sum_{l\in \mathbb{Z}}\sum_{n=0,1,2} t_n \phi^{*}(\omega,\vb{q},l) \phi(\omega,\vb{q},l+n)
    +\tilde{\lambda} \sum_{i=\pm 1}\sum_{j=\pm 1} \phi^*(\omega,\vb{q},i)\phi(\omega,\vb{q},j),
\end{align}
where
\begin{align}
    \tilde{\lambda}=-\frac{\lambda^2 k_F^2 q^2}{32\pi^3} \frac{1}{\frac{\omega^2}{v_{\Phi}^2}-q^2-k_{\Phi}^2}.
\end{align}
Intuitively, the first term—originating from the delta function component—resembles a lattice vibration problem with interactions extending to next-nearest neighbors. Meanwhile, the effective density-density interaction can be interpreted as representing impurities located at sites $\pm 1$. We can write $S_{\textrm{eff}}=\vec{\phi}^{\dagger} M \vec{\phi}$, where $\vec{\phi}$ is indexed by $l$ and $M$ is a real symmetric matrix. We diagonalize $M$ by looking for normal modes: $\sum_b M_{ab} u_b=\xi u_a$.

Without impurities, the eigenvalues of the matrix $M^{\prime}$ associated with $t_n$ entries only are $\xi_k=t_0+2t_1 \cos k+2t_2 \cos 2k$. Note the impurities do not have any effect on the antisymmetric eigenfunction. For symmetric solutions, we first write the eigenvalue equation as
\begin{align}
    \sum_b \kh{\xi-M'_{ab}} u_b=\xi u_a-\sum_n t_n u_{a+n}=\sum_{b} c_{ab} u_b,
    \label{Diag_Eq_EigenFunction1}
\end{align}
where the only nonzero coefficients of the matrix $c$ are
\begin{align}
    c_{1,1}=c_{1,-1}=c_{-1,1}=c_{-1,-1}=\lambda^{\prime}.
\end{align}
Then we can find the inverse of the matrix $(\xi-M')$ as follows:
\begin{align}
    \kh{(\xi-M')^{-1}}_{ab}=\int\frac{\diff{k}}{2\pi}\frac{1}{\xi-\xi_k} \euler^{\ii k(a-b)}\equiv h(a-b).
\end{align}
Multiplying both sides of Eq.~\eqref{Diag_Eq_EigenFunction1} by $h$, we have
\begin{align}
    u_a=\sum_{b,n} h(a-b) c_{bn} u_n.
\end{align}
For this equation to have nonzero solutions, the eigenvalues $\xi$ must satisfy
\begin{align}
    h(0)+h(2)=\frac{1}{2 \tilde{\lambda}}\label{NFL_Eq_Det}.
\end{align}

To extract the spectrum, we simply set $\xi=0$ in Eq.~\eqref{NFL_Eq_Det}. The equation now takes the following form:
\begin{align}
    \int_0^{2\pi}\frac{\diff{k}}{2\pi} \frac{\frac{\omega}{v_F q}}{\frac{\omega}{v_F q}-\cos k}=
    \frac{2\pi v_F}{k_F \lambda^2}\fkh{\frac{\omega^2}{v_{\Phi}^2}-q^2+(k_c^2-k_{\Phi}^2)},\label{NFL_Eq_Dispersion}
\end{align}
where $k_{c}^2=\frac{k_F}{2\pi v_F} \lambda^2$. We find the same particle-hole continuum for $\omega\in [-v_F q,v_F q ]$. This is similar to the situation in Sec.~\ref{Sec_BC_IntF}, where the energy shift of the particle-hole continuum disappears when we let the number of patches $N$ go to infinity.

Eq.~\eqref{NFL_Eq_Dispersion} also has collective mode solutions. From now on, we fix $\lambda$ and consider varying $k_\Phi^2$. Evaluating the integral in Eq.~\eqref{NFL_Eq_Dispersion}, $s=\frac{\omega}{v_Fq}$ satisfies
\begin{align}
    \frac{s}{\sqrt{s^2-1}}=\frac{1}{k_c^2}\kh{s^2 \frac{v_F^2}{v_{\Phi}^2}q^2-q^2+k_c^2-k_{\Phi}^2},
\end{align}
under the condition $\abs{s-\sqrt{s^2-1}}<1$. When $\abs{s-\sqrt{s^2-1}}>1$, the integral becomes $-\frac{s}{\sqrt{s^2-1}}$ and the solution is exactly the negative of the solution to the previous equation.

It is not hard to show that the function $\frac{s}{\sqrt{s^2-1}}$ always has an intersection with the quadratic function on the right-hand side. So there is always a real energy collective mode solution with $s>1$. For $k_{\Phi}^2\geq k_c^2$, this is the only possibility. However, for $k_{\Phi}^2< k_c^2$, there is also a pure imaginary solution that satisfies the condition on $s$. Set $s=\ii \eta$, and then
\begin{align}
    \frac{\eta^2}{\sqrt{\eta^2+1}}=\frac{1}{k_c^2}\kh{-\eta^2 \frac{v_F^2}{v_{\Phi}^2}q^2-q^2+k_c^2-k_{\Phi}^2}.
\end{align}
By the same argument, for $q$ that satisfies $q^2<k_c^2-k_{\Phi}^2$, we will have a solution $\eta>0$. The appearance of the imaginary solution signals the instability in the system. Equivalently, the Hamiltonian corresponding to Eq.~\eqref{NFL_Eq_Seff} is not positive-definite, which we directly verify using the finite-patch approximation.

For $k_c^2=k_\Phi^2$, we find $s\simeq\frac{k_cv_\Phi}{v_Fq}$ as $q\rightarrow 0$, so $\omega(q)\simeq k_cv_\Phi$. This means the collective mode is gapped for small $q$. As a result, the contribution of the collective mode to the structure factor $G(q)$ is negligible in this system, in contrast to the case discussed in Sec.~\ref{Sec_BC_IntF}.

\section{Zero sound in Landau Fermi liquid theory}\label{App_ZeroSound}
In this section, we give a quick derivation of the collective mode from the kinetic equation in the Landau Fermi liquid theory~\cite{coleman2015introduction}. We will denote the quasiparticle distribution function as $n_{\vb{p}}(t,\vb{x})$.
For small fluctuations of the Fermi surface, we have the semiclassical Boltzmann equation:
\begin{align}
    \frac{\diff n_{\vb{p}}}{\diff t}&=\frac{\partial n_{\vb{p}}}{\partial t}+\dot{\vb{x}}\cdot\nabla_{\vb{x}}n_{\vb{p}}+\dot{\vb{p}}\cdot\nabla_{\vb{p}}n_{\vb{p}}=0\\
    &=\frac{\partial n_{\vb{p}}}{\partial t}+\nabla_{\vb{p}}\epsilon_{\vb{p}} \cdot\nabla_{\vb{x}}n_{\vb{p}}-\nabla_{\vb{x}}\epsilon_{\vb{p}}\cdot\nabla_{\vb{p}}n_{\vb{p}},\label{ZS_Eq_Boltzmann}
\end{align}
where we used the Hamilton equation in the second line. $\epsilon_{\vb{p}}$ is the quasiparticle energy, which can be expressed using the bare electron energy $\epsilon_{\vb{p}}^{(0)}$ and Landau parameters $f_{\vb{p},\vb{p}^{\prime}}$ as
\begin{align}
    \epsilon_{\vb{p}}=\epsilon_{\vb{p}}^{(0)}+\sum_{\vb{p}^{\prime}} f_{\vb{p},\vb{p}^{\prime}} \delta n_{\vb{p}^{\prime}}.
\end{align}

Consider a monochromatic oscillation with amplitude $A_{\vb{p}}(\omega,\vb{q})$:
\begin{align}
    n_{\vb{p}}(\vb{x},t)=n_{\vb{p}}^{(0)}+A_{\vb{p}}(\omega,\vb{q}) \euler^{\ii(\vb{q}\cdot \vb{x}-\omega t)}.
\end{align}
We can approximate Eq.~\eqref{ZS_Eq_Boltzmann} as
\begin{align}
    -\ii(\omega-\vb{q}\cdot \vb{v_p}) A_{\vb{p}} \euler^{\ii(\vb{q}\cdot \vb{x}-\omega t)}=\ii\vb{q}\euler^{\ii(\vb{q}\cdot \vb{x}-\omega t)}\sum_{\vb{p}^{\prime}} f_{\vb{p},\vb{p}^{\prime}} A_{\vb{p}^{\prime}}
    \cdot \frac{\partial n_{\vb{p}}^{(0)}}{\partial \epsilon_{\vb{p}}} \vb{v_p}.
\end{align}
This is a consistency equation of $A_{\vb{p}}$. It can be simplified further by defining 
\begin{align}
    A_{\vb{p}}=-\frac{\partial n_{\vb{p}}^{(0)}}{\partial \epsilon_{\vb{p}}}\alpha_{\vb{p}},
\end{align}
and then we have
\begin{align}
    (\omega-\vb{q}\cdot \vb{v_p}) \alpha_{\vb{p}} &=\vb{q}\cdot \vb{v_p}\sum_{\vb{p}^{\prime}} f_{\vb{p},\vb{p}^{\prime}} \kh{-\frac{\partial n_{\vb{p}^{\prime}}^{(0)}}{\partial \epsilon_{\vb{p}^{\prime}}}} \alpha_{\vb{p}^{\prime}}\\
    &=\vb{q}\cdot \vb{v_p}\sum_{\vb{p}^{\prime}\in\textrm{FS}} N(0) f_{\vb{p},\vb{p}^{\prime}} \alpha_{\vb{p}^{\prime}}.
\end{align}

We now focus on the special case of interest in which the only non-zero Landau parameter is $F_0$. Denoting the angle between $\vb{p}$ and $\vb{q}$ by $\theta_{\vb{p}}$, we have
\begin{align}
    \alpha_{\vb{p}}=\frac{\cos\theta_{\vb{p}}}{\frac{\omega}{q v_F}-\cos\theta_{\vb{p}}}\int_0^{2\pi}\frac{\diff \theta_{\vb{p}^{\prime}}}{2\pi} F_0 \alpha_{\vb{p}^{\prime}}.
\end{align}
The solution of this equation is $\alpha_{\vb{p}}\propto \frac{\cos\theta_{\vb{p}}}{s-\cos\theta_{\vb{p}}}$, where $s=\frac{\omega}{q v_F}$ satisfies
\begin{align}
    1=F_0 \int_0^{2\pi}\frac{\diff \theta_{\vb{p}^{\prime}}}{2\pi} \frac{\cos\theta_{\vb{p}^{\prime}}}{s-\cos\theta_{\vb{p}^{\prime}}},
\end{align}
which is essentially the same equation as Eq.~\eqref{Diag_Eq_S0Int} if we identify $F_0=2g$. The zero sound velocity is
\begin{align}
    v_0=v_F \frac{1+F_0}{\sqrt{1+2F_0}}.
\end{align}

\end{document}